\documentclass[groupedaddress,reprint,amsmath,amssymb,aps,floatfix,prl]{revtex4-2}

\usepackage{amsmath,gensymb,textcomp,bm,dcolumn,eurosym,array,tabu,multirow,nicefrac,color,graphicx,upgreek}
\usepackage[colorlinks, linkcolor=blue, citecolor=blue, urlcolor=blue, breaklinks=true]{hyperref}

\usepackage{mathpazo,times} 

\newcommand{\ket}[1]{\lvert #1 \rangle}

\usepackage[subsectionbib]{bibunits}
\usepackage{graphicx}
\usepackage{dcolumn}
\usepackage{bm,MnSymbol}
\usepackage{units}
\usepackage{psfrag}
\usepackage{float}
\usepackage{color}

\usepackage{xr}

\usepackage{subfig}
\begin{document}
	\title{Chiroptical effect induced by gravitational waves}
	\author{Haorong Wu}
	\affiliation{School of Physics and Technology, Wuhan University, Wuhan 430072, China}
	\author{Xilong Fan}
	\email{xilong.fan@whu.edu.cn}
	\affiliation{School of Physics and Technology, Wuhan University, Wuhan 430072, China}
	
	\begin{abstract}		
		We propose the gravitational analog of the chiroptical effect for the first time, demonstrating that gravitational waves (GWs) can induce a reversal of photon chirality through the exchange of angular momentum, namely the spin-2-gravitation chiroptical effect. By analyzing the interaction between photon spin angular momentum (SAM) and GWs, we derive the selection rules governing this exchange, which are strictly dictated by the spin-1 and spin-2 nature of the electromagnetic and gravitational fields, respectively. We find that the gravitational chiroptical effect reflects the local nature of SAM which prevents the accumulation of gravitational perturbations over spatial phase windings, and offers a theoretically rigorous tool to probe the chiral structure of GWs. This mechanism provides a novel observational pathway to constrain modified gravity theories, measure the asymmetric properties of compact binaries, and explore parity-violating physics in the early universe.
	\end{abstract}
	\maketitle

	\indent \emph{Introduction.}\rule[2pt]{8pt}{1pt}
	Chirality, fundamentally defined by the geometric breaking of mirror symmetry, is a ubiquitous concept with profound implications across many physical sciences \cite{PhysRevLett.134.103801,liu2023detection,schilthuizen2005convoluted,cook1903spirals}. Ranging from the fundamental parity violation that dictated the symmetry breaking of the early universe \cite{ym2n-lzts,bcz6-xxn8,PhysRevLett.123.031305,PhysRevLett.102.231301} to homochirality, which serves as the chiral recognition mechanism in biological systems \cite{ozturk2022origins}, chirality governs structure and function at every scale. At the molecular scale, it plays a crucial role in technologies \cite{PhysRevLett.126.123001} extending from drug design \cite{eastgate2017design,patel2019pharmaceuticals} and enantioselective synthesis \cite{pan2019highly,melchiorre2008asymmetric} to emerging fields like molecular spintronics \cite{suda2019light,gohler2011spin}, particularly through phenomena such as the chirality-induced spin selectivity effect \cite{latawiec2025detecting}. In optics, chirality is intrinsically related to the helicity of the electromagnetic (EM) field, manifesting as circular polarization or photon spin angular momentum (SAM) \cite{PhysRevLett.104.163901,PhysRevLett.121.043901}. Beyond standard plane waves, chiral properties have recently been observed in complex optical topologies, such as toroidal spatiotemporal optical vortices, which exhibit transverse orbital angular momentum (OAM) \cite{PhysRevLett.132.153801}. In quantum optics, chiroptical effects emerge when light-matter interactions depend on the chirality of both the light field and the matter \cite{PhysRevLett.122.103201}. These interactions manifest not only as spectroscopic signatures, including circular dichroism \cite{PhysRevLett.123.066803}, optical rotatory dispersion \cite{du2024simple}, and circular differential scattering \cite{PhysRevLett.122.103201}, but also as mechanical effects, such as enantioselective optical forces capable of sorting chiral matter \cite{PhysRevA.91.053824}.

	The concept of chirality is not limited to EM or matter fields but extends naturally to gravitational fields, where gravitational waves (GWs) can exhibit handedness. Chiral GWs are characterized by a statistical or amplitude asymmetry between left- and right-handed circular polarizations \cite{PhysRevLett.102.231301,PhysRevLett.123.031305}. Theoretically, such signatures may arise from various mechanisms in the early universe, including Chern-Simons coupling between the inflaton and gravity \cite{PhysRevLett.83.1506}, gravity at a Lifshitz point \cite{PhysRevLett.102.231301}, the presence of gauge fields during inflation \cite{aoki2025effective}, and specific graviton self-couplings \cite{PhysRevD.85.123537}. Beyond these primordial sources, chirality manifests in the late universe through astrophysical processes. Compact binary coalescences, for instance, emit GWs with circular polarization that depends on the inclination of the system OAM relative to the line of sight \cite{PhysRevLett.124.211301}. Moreover, generic metric theories of gravity extend beyond General Relativity by permitting up to four additional polarization states: the vector $x$ and $y$ modes (helicity $\pm 1$) and the scalar breathing and longitudinal modes (helicity 0) \cite{PhysRevD.110.064073}. The detection of any vector or scalar mode would present a direct contradiction to General Relativity, providing clear evidence of physics exceeding Einstein’s theory of gravity \cite{PhysRevLett.120.201102,PhysRevLett.120.031104}. Consequently, probing the chiral structure of GWs offers a powerful observational tool to investigate parity-violating phenomena in the early universe, measure the asymmetric properties of compact binary systems, and constrain modified theories of gravity.
	
	The interaction between photons and GWs has been studied extensively. For instance, it has been demonstrated that GWs can induce a rotation of the polarization plane in light fields \cite{cruise1983interaction}. Beyond the limit of geometric optics, solutions have been derived for the propagation of free EM radiation within a plane-wave GW background \cite{EnricoMontanari_1998}. Furthermore, researchers have proposed detecting anisotropic GW backgrounds, based on the radiative transfer of photon polarization resulting from forward scattering with gravitons \cite{PhysRevD.98.023518}. Additional studies suggest that photon-graviton conversion may occur when EM fields or GWs propagate through strong magnetic fields \cite{fujita2020gravitational,PhysRevD.107.125027}. More recently, the influence of GW backgrounds on the coherence and high-dimensional entanglement of OAM in twisted photons traversing curved spacetime has been explored \cite{PhysRevD.106.045023}. Our previous work examined the interaction between photon OAM and GWs \cite{q38s-k2tq}. We found that for a photon with an initial OAM state $l$, the interaction with a GW can excite sideband modes $l \pm 1$ and $l \pm 2$ with transition probabilities of approximately $P_{l\pm1} \sim 10^{-17}$ and $P_{l\pm2} \sim 10^{-20}$, respectively. Based on these findings, we proposed a new GW detection technique that offers potentially high sensitivity across a broad frequency spectrum \cite{q38s-k2tq}.
	
\begin{figure} [tbhp]
		\centering
		\includegraphics[width=1\linewidth]{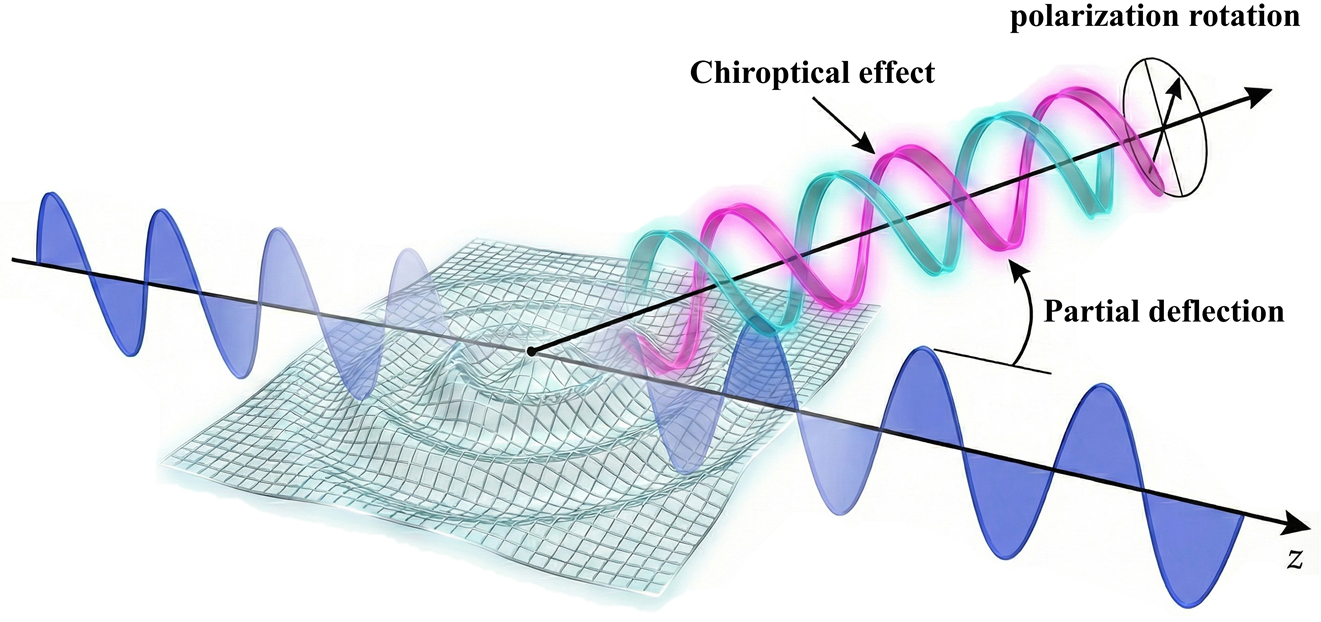}
		\caption{Schematic illustration of three GW effects on the photons: Partial deflection, polarization rotation, and the chiroptical effect.}
		\label{fig.ThreeEffects}
	\end{figure}

	In this study, we investigate the propagation of photons in the presence of GWs and deduce the resulting gravitational effects on them, as shown in Fig. \ref{fig.ThreeEffects}. Our primary focus is the gravitational analog of the chiroptical effect. We identify and quantify the exchange of angular momentum (AM) between the photons and the GWs, and interpret these results within the framework of an effective theory for quantum gravity. Crucially, because SAM is a local property, the gravitational chiroptical effect is predicted to be extremely weak compared to the interaction between photon OAM and GWs \cite{q38s-k2tq}. Nevertheless, this effect may offer a promising, albeit challenging, method to probe the chiral structure of GWs when combined with advanced quantum precision measurement techniques \cite{de2019gravitational}. Unless otherwise specified, units with $c=G=\hbar=1$ are used.  All Greek indices run over $\{0, 1, 2, 3\}$, whereas Latin indices run over $\{1, 2, 3\}$.
	
	\indent \emph{Photon states in GWs.}\rule[2pt]{8pt}{1pt} Without loss of generality, let us consider a monochromatic light with a wave vector $k\hat z$ and a polarization chirality $\sigma$ propagating along the $z$ axis that begins interacting with a monochromatic GW at time $t=0$. The GW propagates at an angle $\theta$ with respect to the $z$ axis, situated in the $xz$ plane. Note that the right- (left-) handed circular polarization with $\sigma=-1~(1)$ has a SAM of $-1~(1)$, with respect to the propagation direction \cite{suppleSAM}. We employ the Dirac notation $\ket{k_1,k_2,k_3,\sigma}$ to denote a photon state with a wave vector $k_1 \hat x+k_2 \hat y+k_3 \hat z$ and a polarization chirality $\sigma$. Therefore, the initial photon state can be written as $\ket {0,0,k,\sigma}$. By invoking perturbation theory, one can find that this state evolves in GWs as \cite{suppleSAM}, to first-order of GW amplitudes,  \begin{align}
		\ket {0,0,k,\sigma}\rightarrow& \ket {0,0,k,\sigma}+D_{1,\sigma} \ket {k_g\sin\theta,0,k+k_g\cos\theta,\sigma}\nonumber \\&+D_{2,\sigma} \ket {k_g\sin\theta,0,k+k_g\cos\theta,-\sigma},\label{eq.StateInGWs}
	\end{align}
	where \begin{align}
		D_{1,\sigma}=& f(t)\frac{\cos^2(\theta/2)}{2}\left [A_+\left (3\cos\theta-1+\frac {2k}{k_g}\right )-2i\sigma A_\times \right ] ,\\
		D_{2,\sigma}=& f(t) \frac {4iA_\times \sigma \cos \theta+A_+(3+\cos 2\theta)}{8},\\
		f(t)=&1-e^{-2ik_g t\sin^2(\theta/2)},
	\end{align}
	$A_+$ and $A_\times$ denotes the GW amplitude for $+$ and $\times$ polarization, respectively, and $k_g$ is the GW frequency. The state on the right-hand side of Eq. \eqref{eq.StateInGWs} is unnormalized. However, since the coefficients $D_{1,\sigma}$ and $D_{2,\sigma}$ are proportional to the extremely small GW amplitudes $A_\lambda$, with $\lambda=\{+,\times\}$, the normalization factor remains approximately unity and is therefore neglected.
	
	Several consequences of the GW interaction, including the gravitational chiroptical effect, can be deduced from the state \eqref{eq.StateInGWs}. These phenomena are better elucidated by expanding the GWs in the circular polarization basis. Analogous to the EM fields, we can define the circular polarization amplitudes for GWs by expanding $A_+=(A_R+A_L)/\sqrt 2$ and $A_\times=i(-A_R+A_L)/\sqrt 2$ \cite{misner1973gravitation}, where $A_R$ and $A_L$ are the amplitudes for the right and left circular polarizations, respectively. For simplicity, we will use $A_R$ or $A_L$ to represent GWs with right or left circular polarizations, respectively. 
	
	For a monochromatic GW propagating along the $z$ axis, described by $h_{\mu\nu}=(A_Re^R_{\mu\nu}+A_Le^L_{\mu\nu})e^{ik_g(-t+z)}$, with $e^R_{\mu\nu}$ and $e^L_{\mu\nu}$ being the right and left circular polarization basis for GWs, respectively, the (averaged) energy density in the transverse-traceless (TT) gauge is given by $T_{00}=k^2_g\big ( \left | A_R\right |^2+\left | A_L\right |^2 \big )/(32\pi)$ \cite{misner1973gravitation}. It is instructive to interpret these results within the framework of gravitons. While we do not claim to demonstrate the existence of gravitons, employing this concept facilitates a clearer physical picture of the result. In the context of the effective theory for quantum gravity, if the GWs can be quantized as gravitons with energy $k_g$ \cite{donoghue1994general,lagouvardos2021gravitational,PhysRevD.109.044009}, one can calculate the number density of gravitons as \cite{dyson2013graviton,PhysRevD.109.044009} $N=T_{00}/k_g=k_g\big ( \left | A_R\right |^2+\left | A_L\right |^2 \big )/(32\pi)$. Further, the (average) AM density of the GW is given by \cite{RevModPhys.52.299,xin2021spin} $s^{\rm GW}_i= \left <\epsilon_{ijk}{\rm Re}(h_{lj}){\rm Re}(h_{kl})_{,0} \right >/(16\pi)$, where $\epsilon_{ijk}$ is the Levi-Civita symbol. The only nonvanishing component is $s^{\rm GW}_3=k_g\big ( -\left | A_R\right |^2+\left | A_L\right |^2 \big )/(16\pi)$. Analogous to photons, $s^{\rm GW}_i$ corresponds to the SAM density of gravitons. For $A_R$, the SAM per graviton is $s^{\rm GW}_3/N=-2$; similarly, the SAM per graviton in $A_L$ is $2$. These results extend naturally to GWs propagating in an arbitrary direction: The right-handed component carries an AM density of $- k_g \left | A_R\right |^2/(16\pi)$, while the left-handed component carries a density of $k_g \left | A_L\right |^2/(16\pi)$, measured with respect to the propagation direction.
	
	\indent \emph{Partial deflection and polarization rotation.}\rule[2pt]{8pt}{1pt} Before exploring the chiroptical effect, we would like to briefly discuss the other two effects.
	The first effect induced by GWs is partial deflection. After interacting with GWs, the photon acquires two new states, whose wave vectors are given by $\mathbf k'=k_g\sin\theta\hat x+(k+k_g\cos\theta)\hat z$. This wave vector forms an angle $\phi_{\rm GW}=\arctan(k_g\sin\theta/(k+k_g\cos\theta))$ with respect to the original wave vector $\mathbf k=k\hat z$. In most circumstances, however, $k_g$ is far less than $k$. For example, $k_g\sim 10^{-5}~{\rm m}^{-1}$ for kHz GWs and $k\sim 10^7~{\rm m}^{-1}$ for visible light. Then, to first order of $k_g/k$, $\phi_{\rm GW}\approx k_g \sin\theta/k\sim 10^{-12}\sin\theta$, which is too small to be significant. Therefore, we will neglect the deflection of the photon wave vector in GWs, and henceforth, we will focus on the polarization states and write the photon states in GWs as\begin{equation}
		\ket {\sigma}\rightarrow (1+D_{1,\sigma})\ket {\sigma}+D_{2,\sigma} \ket{-\sigma},\label{eq.StateInGWs2}
	\end{equation}
	where $\ket {\pm \sigma}=\ket {0,0,k,\pm\sigma}$. 
	
\begin{figure} [tbhp]
		\centering
		\includegraphics[width=1\linewidth]{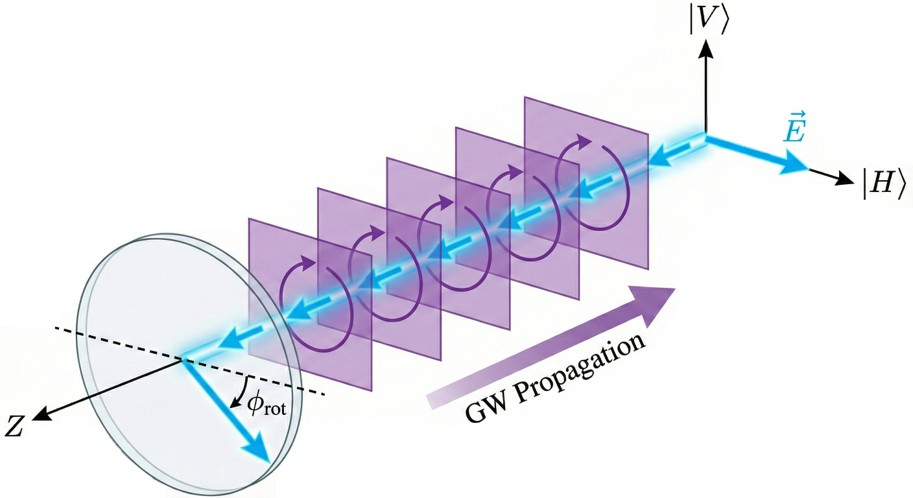}
		\caption{A right-circularly polarized GW propagating along the $-z$ axis rotates the light polarization clockwise. Note that the direction of rotation matches the chirality of the GW.}
		\label{fig.PolarizationRotation}
	\end{figure}

	The second effect is polarization rotation. The linear and circular polarization states of photons are related by $\ket H=(\ket {-1}+\ket{1})/\sqrt 2$, $\ket V=i(\ket {-1}-\ket{1})/\sqrt 2$, where $\ket H$ and $\ket V$ are the horizontal and vertical linear polarization states, respectively, and $\ket {-1}$ and $\ket {1}$ are the right- and left-circular polarization, respectively. Suppose, initially, the photons are horizontally polarized. After interacting with GWs, the photon state is given by $\ket H \rightarrow C_H\ket H+C_V\ket V$, where $C_H=1+ f(t)A_+k(1+\cos \theta)/(2 k_g)$, $C_V=  f(t)A_\times (1+2\cos \theta)/2$, and only leading-order terms in $k_g$ are retained. The state has a rotation angle $\Delta \phi$ given by\begin{align}
		\phi_{\rm rot}=&\arctan\frac{{\rm Re}(C_V)}{{\rm Re}(C_H)}\approx A_\times (1+2\cos\theta) \sin^2\left (k_g t\sin^2 \frac{\theta}2 \right )\nonumber \\
		=&\frac {(A_R-A_L)}{2\sqrt 2}(1+2\cos\theta)\sin\left (2k_g t\sin^2 \frac{\theta}2\right ).\label{eq.polrot}
	\end{align}
	Our result is consistent with previous studies \cite{cooperstock1993laser,EnricoMontanari_1998,de2019gravitational} which demonstrate that only $A_\times$ induces polarization rotation of light. For a special case where $A_R$ propagates along the $-z$ direction, the polarization of light will be rotated clockwise by an angle $A_R \sin(2k_g t)/(2\sqrt 2) $, as shown in Fig. \ref{fig.PolarizationRotation}; similarly, $A_L$ will rotate the light polarization counterclockwise by an angle $A_L \sin(2k_g t)/(2\sqrt 2)$. Note that the direction of the rotation matches the polarization direction of the GW.
	
	\indent \emph{Spin-2-gravitation chiroptical effect.}\rule[2pt]{8pt}{1pt}Finally, we examine the gravitational analog of the chiroptical effect. From Eq. \eqref{eq.StateInGWs2}, we find that a photon initially in state $\ket \sigma$ possesses a probability amplitude for chirality reversal given by
	\begin{align}
		S_{\sigma\rightarrow -\sigma}=&f(t)
		[4iA_\times \sigma \cos \theta+A_+(3+\cos 2\theta)]/8\nonumber \\
		=&f(t)
		[(\cos\theta-\sigma)^2 A_L+(\cos\theta+\sigma)^2A_R]/(4\sqrt 2).\label{eq.chiro}
	\end{align}
	The gravitational chiroptical effect is different from the polarization rotation, as $A_+$ does not cause the latter effect, whereas it actively contributes to the chiroptical effect.
	
\begin{figure*}[tbhp]
		\centering
		\subfloat[][quasi-parallel]{%
			\includegraphics[height=0.27\linewidth]{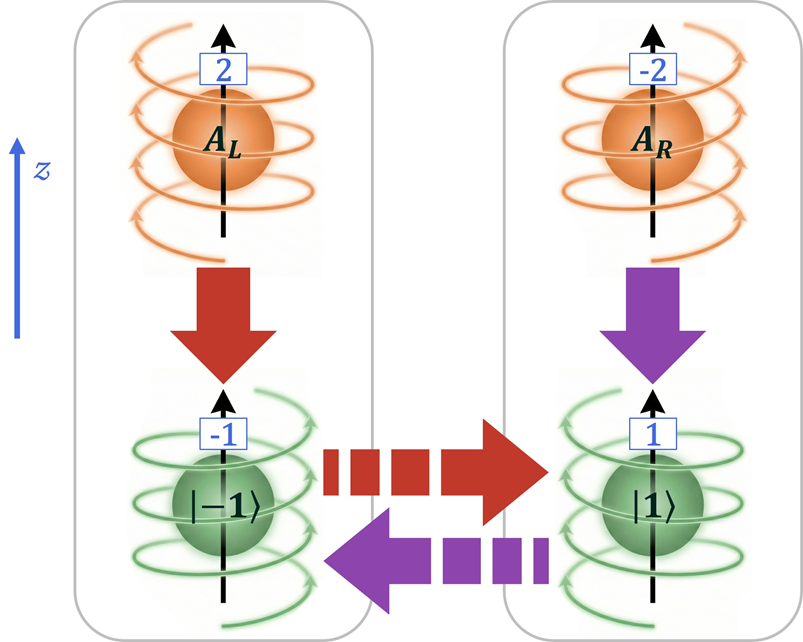}%
			\label{fig.para}
		}\hfill
		\subfloat[][antiparallel]{%
			\includegraphics[height=0.27\linewidth]{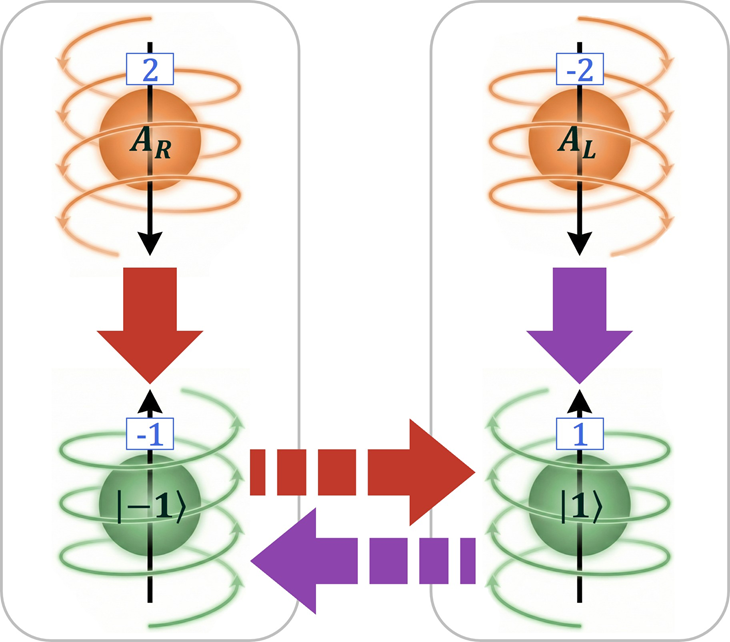}%
			\label{fig.antipara}
		}\hfill
		\subfloat[][perpendicular]{%
			\includegraphics[height=0.27\linewidth]{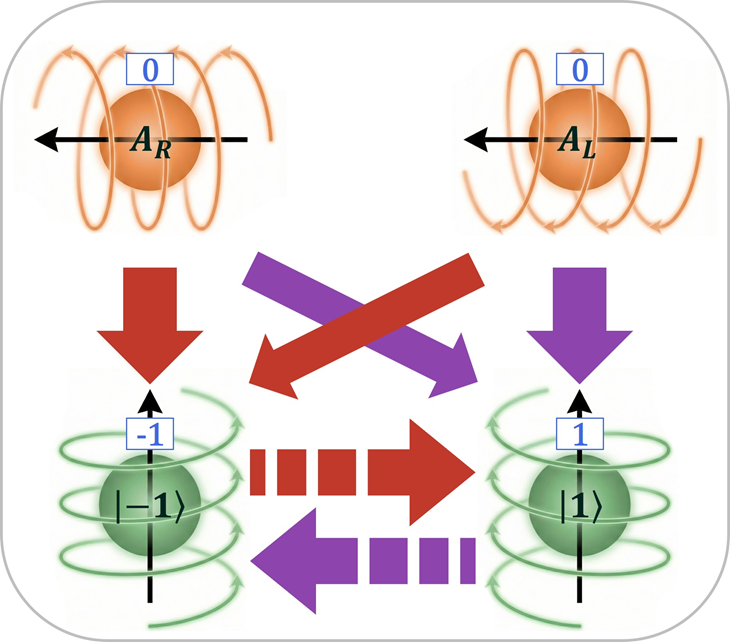}%
			\label{fig.perp}
		}
		\caption{Schematic of the gravitational chiroptical effect during (a) quasi-parallel ($\theta \rightarrow 0$), (b) antiparallel ($\theta = \pi$), and (c) perpendicular ($\theta = \pi/2$) propagation. Orange and green spheres represent GWs (gravitons) and photons, respectively. The black arrows indicate propagation directions and the blue numbers in boxes represent the $z$-component of AM. The red and purple arrows depict the reversal of photon chirality resulting from the exchange of AM between photons and GWs. The diagrams show that the selection rules governing the exchanges are strictly dictated by the spin-1 and spin-2 nature of the electromagnetic and gravitational fields, respectively.
		}
		\label{fig.chiroptic}
	\end{figure*}

	In the gravitational chiroptical effect, the SAM of photons \cite{PhysRevResearch.4.023165,barnett2016natures} is altered. It is natural to ask: What is the origin of this variation? Does the GW provide or absorb the AM? Or is it merely a background facilitating the exchange of AM between photons? These questions may be addressed within the framework of quantum mechanics by assuming that GWs can be quantized in the form of gravitons, and a right- (left-) circular polarization graviton has the SAM of $-2$ ($2$) with respect to its propagation direction, respectively. We will focus on three special cases: quasi-parallel ($\theta\rightarrow 0$), antiparallel ($\theta=\pi$), and perpendicular ($\theta=\pi/2$) propagation. The diagrams for these cases are shown in Fig. \ref{fig.chiroptic}. Orange and green spheres represent GWs (gravitons) and photons, respectively. The black arrows indicate propagation directions and the blue numbers in boxes represent the $z$-component of AM. The red and purple arrows depict the reversal of photon chirality resulting from the exchange of AM between photons and GWs.

	When the GWs and the light propagate parallel to each other with $\theta=0$, the function $f(t)$ will vanish. This is equivalent to stating that when gravitons and photons propagate in the same direction, their velocity difference is zero, precluding any interaction. To manifest their interaction, we can let them propagate in the quasi-parallel limit with $\theta \rightarrow 0$, so that $f(t)\ne 0$ and we can focus on the remaining components of $S_{\sigma\rightarrow -\sigma}$, giving\begin{equation}
		S_{\sigma\rightarrow -\sigma}=\begin{cases}
			2^{-1/2}f(t)A_L, & \sigma=-1, \\
			2^{-1/2}f(t)A_R, & \sigma=1.
		\end{cases}
	\end{equation}
	In this case, $A_R$ cannot interact with right-circular polarized photons ($\sigma=-1$). This may be explained by the fact that the gravitational field is a spin-2 field, and, hence, the gravitons within $A_R$ carry a SAM of -2 (with respect to the $+z$ direction). Therefore, $A_R$ cannot couple to the right-circular polarized photons, which require a transfer of AM to make quantum transition for photons from $\ket {-1}$ to $\ket{1}$ with photon SAM changing from $-1$ to $1$ (the OAM is not considered here). On the other hand, $A_L$ can couple to the right-circular polarized photons, since $A_L$ can provide the AM that the photons need to make such transition. For a general photon polarization state $c_1\ket {1}+c_{-1}\ket {-1}$ interacting with $A_R$, with $|c_1|^2+|c_{-1}|^2=1$, the right-circular part $c_{-1}\ket {-1}$ is invariant, since it cannot couple to $A_R$, while the left-circular part $c_{1}\ket {1}$ has a non-zero probability of reversing its chirality as $c_{1}\ket {1}\rightarrow c_{1}\left [\ket {1}+2^{-1/2}f(t)A_R \ket {-1}\right ]$, where the term containing $D_{1,1} $ is ignored as $D_{1,1} \ll 1$. Thus, we could answer the earlier questions. The AM needed to change the photon chirality should come from GWs, rather than other photons. This can be viewed as a exchange of AM between the photon state $\ket{1}$ and the GW $A_R$. Similarly, the GW $A_L$ is only coupled to the photon state $\ket {-1}$. The AM exchanges $A_L \leftrightarrow \ket{1} $ and $A_R \leftrightarrow \ket{-1}$ are forbidden, as shown in Fig. \ref{fig.para}.
	
	If the GWs propagate along the $-z$ direction with $\theta=\pi$, $A_R$ and $A_L$ will have the AM of $2$ and $-2$ with respect to the $+z$ direction, respectively. The probability amplitude of the chiroptical effect is given by
	\begin{equation}
		S_{\sigma\rightarrow -\sigma}=\begin{cases}
			2^{-1/2}f(t)A_R, & \sigma=-1, \\
			2^{-1/2}f(t)A_L, & \sigma=1.
		\end{cases}
	\end{equation}
	Therefore, AM exchanges are restricted to the channels $A_R \leftrightarrow \ket{-1} $ and $A_L \leftrightarrow\ket{1} $, with other exchanges being forbidden, as shown in Fig. \ref{fig.antipara}. Finally, if the GWs propagate perpendicularly to the light with $\theta=\pi/2$, the probability amplitude is given by\begin{equation}
		S_{\sigma\rightarrow -\sigma}=f(t)(A_L+A_R)/(4\sqrt 2).
	\end{equation}
	This amplitude $S_{\sigma\rightarrow -\sigma}$ is independent of the photon chiralities $\sigma$, and is symmetric about $A_L$ and $A_R$. Thus, we could state that as the $z$ component of GW AM vanishes, the internal structure of GW AM cannot be resolved by the chiroptical effect of photons propagating along the $z$ axis. Consequently, all AM exchange channels between GWs and photons are possible, as shown in Fig. \ref{fig.perp}.

	\indent \emph{Conclusion.}\rule[2pt]{8pt}{1pt}We have examined the propagation of light in the presence of GWs using perturbation theory and identified three different gravitational effects: partial deflection, polarization rotation, and the chiroptical effect. While the latter two are both related to the coefficient $D_{2,\sigma}$ in Eq. \eqref{eq.StateInGWs2}, they are distinct effects. Polarization rotation is induced solely by the $\times$-polarization component ($A_\times$), whereas the chiroptical effect can be induced by both GW polarization amplitudes. Moreover, the chiroptical effect involves AM exchanges between GWs and photons, which are absent in polarization rotation. Certain AM exchanges are forbidden, due to the spin-2 and spin-1 nature of the gravitational and EM fields, respectively.
	
	We estimate the number of photons undergoing chirality reversal to be $N \approx 10^{-23}$ per second under the parameters: GW amplitudes $A_L=A_R=10^{-21}$, GW propagation direction $\theta=\pi/3$, GW frequency $f_g=10 {\rm ~Hz}$, photon chirality $\sigma=-1$, photon wavelength $\lambda=700{\rm ~nm}$, laser power $P=100{\rm W}$, and interaction length $L=5.6\times 10^6 {\rm ~m}$. In comparison, interactions between GWs and photon OAM yield significantly higher signal rates ($N \approx 10^{3}$ for $\ket l \rightarrow \ket{l\pm 1}$ and $N\approx 1$ for $\ket l\rightarrow \ket{l\pm 2}$, where $\ket l$ is the photon state with OAM $l$) \cite{q38s-k2tq}. Hence, the signal arising from the chiroptical effect is extremely small, rendering it currently infeasible for GW detection. This disparity stems from the local nature of SAM ($\sigma$) versus the topological nature of OAM ($l$). OAM cannot be determined by measurements at a single point within the light field. Instead, it is defined by the phase winding around the propagation axis as \cite{allen1992orbital,Gbur:08} $l=\oint_C \nabla  \phi(\mathbf r)\cdot d\mathbf r/(2\pi)$, where $C$ denotes a closed contour encircling the propagation axis, and $\phi(\mathbf r)$ represents the phase of the light field. As a result, OAM is sensitive to the presence of GWs ($h_{\alpha\beta}$) because it accumulates the gravitational perturbation over a spatial region. Conversely, SAM is a local property defined at specific spatial coordinates \cite{PhysRevLett.88.053601}. Its value at a point $x^i$ can be determined via a polarization detector located at that position, requiring neither knowledge of the beam center nor measurements at any other point $x'^i \ne x^i$. SAM is encoded entirely in the temporal behavior of the EM vector at $x^i$. Treating photons as massless particles, we recall that a single particle cannot locally sense the presence of a GW. Indeed, for any single particle, one can construct TT coordinates where the particle appears stationary to first order in $h_{\alpha\beta}$ \cite{flanagan2005basics}. Therefore, observing physical GW effects requires comparing spatially separated points or analyzing gradients. For photons, this non-locality is introduced via terms containing first derivatives of the metric perturbation, $h^{\alpha\beta}_{~~,\mu}$ \cite{suppleSAM}. These terms introduce the GW wave vector $k_g$ as a factor in the equations of motion. As a result, the transition probability for the gravitational chiroptical effect ($\ket \sigma\rightarrow \ket{-\sigma}$) is suppressed by a factor of $(k_g/k)^2$ compared to the OAM case ($\ket l\rightarrow \ket{l\pm 2}$) \cite{q38s-k2tq}. For GWs with frequency $f_g=10 {\rm ~Hz}$ and light with wavelength $\lambda=700{\rm ~nm}$, this suppression factor is $(k_g/k)^2 \approx 10^{-28}$. In summary, under GW interaction, photon SAM behaves as a robust intrinsic property, whereas photon OAM, determined by the spatial phase distribution, is significantly more susceptible to gravitational perturbations.
	
	Despite these challenges, the chiroptical effect may still hold promise for probing the chiral structure of GWs given future technological advancements. By comparing Eqs. (\ref{eq.polrot}-\ref{eq.chiro}), one may find that the probability amplitude of the gravitational chiroptical effect has the same order of magnitude as the angle of polarization rotation. Some have studied the feasibility of utilizing a Hong–Ou–Mandel interferometer to carry out GW detection and spectrometry by measuring the polarization rotation induced by GWs \cite{de2019gravitational}. They find that the measurement could only be carried out with the arm lengths on the order of $10^7 {\rm ~m}$, and by using heralded photon sources of $ 1{\rm ~W}$ power, with frequencies of $10 {\rm ~MHz}$. 
	
	Furthermore, the results of this study may be extended to investigate chiroptical effects induced by GWs with vector (helicity $\pm 1$) or scalar (helicity 0) polarizations, as predicted by alternative metric theories \cite{PhysRevD.110.064073}. Since these additional modes deform spacetime geometry in distinct ways, they are expected to induce distinguishable signatures in photon chirality during propagation. Consequently, isolating the specific chiroptical signatures associated with vector or scalar modes could provide a rigorous test for modified gravity theories \cite{PhysRevLett.120.201102,PhysRevLett.120.031104}.
	
	Moreover, by utilizing only the tensor polarizations of GWs, we demonstrate that the induced chiroptical effect varies with the helicity of the GW circular polarization. We draw an analogy to Brownian motion, where the observable motion of pollen grains implies the existence of invisible atoms. Similarly, the chirality perturbations of photons may provide indirect evidence for the existence of gravitons within the GW background \cite{parikh2020noise,moffat2025stochastic}, notwithstanding current uncertainties regarding their exact interaction mechanisms.
	
	\indent \emph{Acknowledgments.}\rule[2pt]{8pt}{1pt} We thank Lixiang Chen and Yanbei Chen for the helpful and inspiring discussions. This work is supported by National Natural Science Foundation of China (125B2103) and National Key R$\&$D Program of China (2020YFC2201400).
	
	\indent \emph{Data availability.}\rule[2pt]{8pt}{1pt}The data that support the findings of this article are not publicly available. The data are available from the authors upon reasonable request.
	
	
	\IfFileExists{myBib.bib}{
		\bibliography{myBib.bib}   
	}{%
	}
\end{document}